\documentclass[aps,pre,preprint,showkeys,12pt]{revtex4}
\usepackage{graphicx}
\usepackage{psfrag}
\usepackage{fancyhdr}
\usepackage{amssymb}
\begin{document}
\title{Von Neumann Turbulent Transport Model}
\author{J. I. Katz}
\affiliation{Department of Physics and McDonnell Center for the
Space Sciences \\ Washington University, St. Louis, Mo. 63130}
\email{katz@wuphys.wustl.edu}
\date{\today}
\begin{abstract}
I propose a simple model, based on an analogy to von~Neumann artificial
viscosity, of turbulent diffusion, heat diffusion and viscosity
coefficients for use in modeling subgrid turbulent diffusivity in
multi-phase numerical hydrodynamics and, more generally, in subgrid
turbulent viscosity and thermal transport.  In analogy to the von~Neumann
artificial viscosity, these coefficients explicitly contain the grid size
and do not attempt a quantitative model of the unresolved turbulence.
In order to address the problem that it is often not known {\it a priori\/}
when and where a flow will become turbulent, the coefficients are set to
zero when the flow is not expected to be turbulent on the basis of a
Richardson/Rayleigh-Taylor stability criterion, in analogy to 
von~Neumann's setting of artificial viscosity to zero in expanding flows.
\end{abstract}
\maketitle
\section{Introduction}
Turbulent diffusion in heterogeneous or multi-phase flows and turbulent
viscosity and heat transport in all flows are problems of great practical
importance in numerical hydrodynamics.  In many problems the Reynolds number
is enormous and if the flow becomes turbulent it is not possible to
calculate it explicitly on the finer scales of heterogeneity that are still
large enough that microscopic transport processes (diffusivity, viscosity 
and conductivity) have not smoothed out the variations in the corresponding
physical variables.

When the composition of the fluid is heterogeneous the most important
questions usually concern the mixing of material of different compositions
or in different phases \cite{D05}, rather than the turbulent momentum
transfer.  A number of sophisticated turbulence models have been developed,
including the ``K-L'' model \cite{DT06} of Rayleigh-Taylor and
Richtmyer-Meshkov instability and the more general BHRZ model \cite{BHRZ92}
(sometimes called the BHR model) that implicitly account for all sources
driving instability and turbulence, including shear (Kelvin-Helmholtz
instability), through their derivation from the Euler equations.  As
powerful as these models are, they can be complex to implement because they
describe the strength of turbulence with parameters that are non-local and
history-dependent functions of space and time.  The success of the very
simple von~Neumann \cite{vNR50} artificial viscosity, nearly universally
used in numerical calculations of compressible flow, suggests that a
similarly simple model of turbulent transport, may be useful.
\section{General Considerations}
The elementary Kolmogorov turbulence model indicates an eddy turnover time
\begin{equation}
t_k \propto k^{-2/3},
\end{equation}
for eddy size $1/k$.  Hence if the flow is turbulent at all small eddies
turn over much faster than larger ones.  Assuming at least one turnover for
the largest driving eddies there is likely to be rapid mixing down to the
inner turbulence scale.

If the Schmidt number
\begin{equation}
{\rm Sc} \equiv {\nu \over D},
\end{equation}
where $\nu$ is the kinematic viscosity and $D$ the diffusivity, is of order
unity or less then diffusive mixing on scales smaller than the inner
turbulence scale will be at least as rapid as the turnover of the smallest
eddies, and much more rapid than that of the driving eddies.  This is
usually the case because in dilute gases or weakly coupled liquids momentum,
thermal energy and composition are, to a substantial extent, carried by the
same particles (in a plasma momentum is almost entirely carried by the ions
and energy by the electrons; at sufficiently high Reynolds number all of 
these quantities diffuse on the inner turbulence scale much faster than the
turnover of the driving eddies).  Only if ${\rm Sc} \gg 1$ may diffusion
down to the molecular or atomic scale be slower that turbulent turnover;
this may occur when the viscosity is large because of strong intermolecular
forces, as in viscous liquids (glycerin, honey, pitch, the Earth's mantle,
{\it etc.\/}) in which momentum diffuses rapidly but chemical composition
slowly.  Such substances rarely undergo turbulent high Reynolds number
flows, though such flows are, in principle, possible if driven strongly
enough.

In the classical Kolmogorov turbulent cascade the inner scale $\ell_{in}$
of turbulence is approximated by
\begin{equation}
\ell_{in} \approx \left({\nu^3 \over \epsilon}\right)^{1/4},
\end{equation}
where $\nu$ is the kinematic viscosity and $\epsilon \sim U^3/R$ is the
kinematic (per unit density) energy dissipation per unit volume of a flow
with characteristic velocity $U$ on an outer scale $R$.  This leads to an
estimate of the time for microscopic diffusion to produce atomic-scale
homogeneity
\begin{equation}
t_{diff} \approx {\ell_{in}^2 \over D} \approx Sc \sqrt{\nu R \over U^3}
\approx {Sc \over \sqrt{Re}}{1 \over t_{turn}},
\end{equation}
where the turnover time $t_{turn} \equiv R/U$.  In high Reynolds number
flows microscopic (diffusive) mixing on length scales from $\ell_{in}$ down
to atomic dimensions is a very rapid process unless the $Sc \gg 1$.

Turbulent heat transport presents similar problems.  If the Prandtl number
\begin{equation}
{\rm Pr} \equiv {\nu \over \alpha},
\end{equation}
where $\alpha$ is the microscopic heat diffusion coefficient, is of order
unity or less then diffusive heat transfer on scales smaller than the
inner turbulence scale will be at least as rapid as the turnover of the
smallest eddies, and much more rapid than that of the driving eddies.  This
is usually the case because in dilute gases or weakly coupled liquids
momentum and thermal energy are, to a substantial extent, carried by the
same particles.  In liquid metals or ionized plasmas the rapid thermal
conduction by electrons or rapid radiative transfer lead to ${\rm Pr} \ll 1$
and this inequality holds even more strongly.  Just as for turbulent
mixing, only in viscous liquids in which strong intermolecular forces carry
momentum is ${\rm Pr} \gg 1$ and the efficacy of diffusive heat transport on
scales smaller than the inner turbulence scale is problematic; this paper
does not discuss that regime.
\section{Turbulence Criterion}
The problem is to model the subgrid diffusion in fluids with ${\rm Sc}
\lesssim 1$ when small features of the flow cannot feasibly be resolved
computationally.  This requires a criterion to decide when the flow is
turbulent as well as a model for turbulent mix.  It is often not possible to
perform an instability analysis because of the complexity and non-stationary
character of the large scale flow in problems such as inertial confinement
fusion.  The recognition that a complex flow is turbulent and will possess a
turbulent cascade to large wave numbers, much less the determination of its
characteristics, is non-trivial.  This limits the usefulness of even the
most sophisticated turbulence models.  

Because microscopic diffusion on scales smaller than $\ell_{in}$ is
generally much faster than the lifetime of the flow, which is generally
${\cal O}(t_{turn})$, determining whether a flow is turbulent on scales
$\sim \ell_{in}$ not explicitly resolved in a numerical calculation is much
more important than a quantitative estimate of the turbulent diffusion
coefficient itself.  A common criterion for whether a computed or observed
flow is turbulent is ``I know it when I see it''.  This is useless in a
large numerical simulation in which the structure of the flow is not known
in advance.  A simple and automatic turbulence criterion is required.
Without knowing if a flow is turbulent or not it is impossible even to
decide if a turbulence model should be used at all, much less to make
reasonable estimates of its parameters.  

The model presented here is almost childishly na\"ive compared to the
non-oscillatory finite volume (NFV) difference schemes\cite{MR02,MRG06}
that have been remarkably successful in modeling turbulent flows.  So is the
von~Neumann artificial viscosity, yet it remains the basis of most numerical
calculations of compressible flows.

The present model uses an on-off switch on
turbulent transport based on the Richardson stability criterion appropriate
to high Reynolds number shear flow (if there is no shear it reverts to the
Rayleigh-Taylor stability criterion for inviscid miscible fluids).  This
switch is a generalization of one used (\cite{DT06}) in the ``K-L'' model
of turbulence produced by Rayleigh-Taylor and Richtmyer-Meshkov instability.
An implicit switch is present in the more general BHRZ model \cite{BHRZ92}
of turbulence in which the explicitly calculated velocity field drives or 
damps turbulence according to the Euler equations.

A switch is not needed in calculations of flows (such as unstratified pipe
flows) that are known {\it a priori\/} to be unstable, and that constitute
the usual tests of turbulence models.  It is essential in more complex 
heterogeneous flows that contain both stable and unstable regions that are
not predictable in advance.  The switch is analogous to the switch in the
von~Neumann artificial viscosity that sets it to zero in rarefying regions.

Sub grid scale (SGS) turbulence models of this type were first proposed by
Smagorinsky \cite{S63} for use in numerical large-eddy simulations (LES) of
flows whose small scale turbulence is numerically unresolved.  Many SGS
models have been developed since his pioneering work \cite{LM96,P00,KPS02,
YM07}, but none appear to be universally applicable.  In this paper I follow
von~Neumann's \cite{vNR50} artificial viscosity model for dissipation in
unresolved shocks, by renouncing any attempt to model the SGS turbulence
in detail or on the basis of fundamental principles, just as \cite{vNR50}
renounced any attempt to understand the microscopic dissipation mechanisms
and structure of shocks (which also depend on the detailed physics of the
particular fluids involved, so that no single model can be generally valid),
and construct a model that instead depends explicitly on the numerical
properties of the LES, while incorporating a criterion for instability and
the {\it presence\/} of turbulence and turbulent diffusivity that explicitly
depends on the resolved properties of the flow.
\section{Model Turbulent Diffusivity}
This problem has some similarity to that of modeling shocks in compressible
flow.  In each problem the difficulty is posed by subgrid scale dissipative
processes that cannot be resolved computationally, but are known to be
present.  The classic solution\cite{vNR50} is to define an ``artificial
viscosity'' that is explicitly dependent on the grid size (thus broadening a
shock to a width of a few resolution elements) and introducing a nonlinear
criterion for its presence, in essence an on-off switch that turns it on
when the fluid is being compressed and off when being rarefied (because in
thermodynamically stable systems there are no rarefaction shocks).  The
resulting artificial viscosity is strongly nonlinear and has the unphysical
property of a magnitude that depends on the scale of the computational grid.
It has been widely adopted because it accurately reproduces the shock jump
conditions, at the price of not reproducing the structure of the shock on
the scale of the grid resolution or its (much finer and unresolvable)
actual microscopic structure.  It has the same dimensions as real physical
viscosity.

We seek an analogous heuristic artificial diffusivity for multicomponent flows.
The turbulent diffusivity we propose is
\begin{equation}
D_{turb} = \left\{
\begin{array}{cl}
0 & {\rm if}\ {{\vec a} \cdot \nabla \rho^\prime \over \rho^\prime \vert e_{ij}
\vert^2} \ge {1 \over 4},\\
{\Delta^2 \vert e_{ij} \vert \over 6} & {\rm if}\ {{\vec a} \cdot \nabla
\rho^\prime \over \rho^\prime \vert e_{ij} \vert^2} < {1 \over 4};
\end{array} \right.
\end{equation}
where $\vec a$ is the local fluid acceleration, $\Delta$ is a characteristic
(filter) length, 
\begin{equation}
e_{ij} \equiv {\partial u_i \over \partial x_j} + {\partial u_j \over \partial
x_i} - {2 \over 3} \delta_{ij} {\partial u_k \over \partial x_k},
\end{equation}
is the traceless (anisotropic) symmetrized strain rate tensor \cite{LL59,B67}
and
\begin{equation}
\vert e_{ij} \vert = \sqrt{e_{ij} e_{ij}}
\end{equation}
is its scalar magnitude; summation over repeated indices is implicit.  This
$D_{turb}$ is analogous to the Smagorinsky turbulent viscosity model of
subscale eddy simulation (see \cite{GO93} for a review).

A number of heuristic prescriptions for $\Delta$ have been used
\cite{CM93,F93,P93}.  Without an opportunity to compare to detailed
experimental data in complex and difficult to diagnose flows such as
those encountered in inertial confinement fusion, it is hard to decide
which is most appropriate for those flows, and the best prescription may
vary over space and time.  We suggest that $\Delta$ be taken to be the
smallest dimension of a spatial zone because high aspect ratio zones are
generally used only when there are large gradients of composition or other
variables in the direction of the smallest side of a zone that need to be
resolved; these smallest sides are likely to limit mixing lengths.

Alternative forms are possible.  For example, a tensor $D_{ij}$
could be defined by replacing $\vert e_{ij} \vert$ by $e_{ij}$ in the
expression for $D_{turb}$.  However, if the turbulent cascade isotropizes
the turbulence, as generally assumed in the absence of a quantitative more 
general model, then the subgrid $D_{turb}$ should be a scalar, as in the
form above.

The on-off switch is adapted from the Richardson
instability criterion \cite{T73,DR81} in stratified fluids, replacing the
stratifying acceleration of gravity $\vec g$ by the component of local
acceleration $\vec a$ along the gradient of potential density:
\begin{equation}
\nabla \rho^\prime \equiv \nabla \rho - \left.{\partial \rho \over \partial p}
\right\vert \nabla p,
\end{equation}
where the partial derivative is evaluated under thermodynamic conditions
appropriate to matter displacements.  If flows are nearly adiabatic on
displacement time scales this is $\left.{\partial \rho \over \partial p}
\right\vert_S$, while if they are nearly isothermal (as will be the case if
conductive or radiative energy transfer to a heat bath is rapid) it is
$\left.{\partial \rho \over \partial p} \right\vert_T$.  Stratification can
stabilize against either (or both) of Rayleigh-Taylor or Kelvin-Helmholtz
instabilities.

This criterion ignores the stabilizing effect of viscosity, which is small
at high Reynolds numbers and not easy to calculate when both shear and
stratification affect stability.  At lower Reynolds numbers (defined by zone
size), viscosity is significant but calculated directly by a hydrodynamics
code.  When this is the case subgrid scale turbulence does not occur and,
provided the Schmidt number is ${\cal O}(1)$ molecular diffusivity is rapid
and no turbulent diffusivity model is required.

The factor of $1/6$ is introduced to allow for three spatial dimensions (in 
analogy with the factor $1/3$ in the elementary kinetic theory result $D =
u \ell / 3$, where $\ell$ is the mean free path), and for the double
counting resulting from symmetrization of $e_{ij}$.  Because the proposed
form is phenomenological, if fitting data are available it may be
appropriate to introduce an additional dimensionless multiplicative factor
chosen to match those data.  Such ``data'' may be obtained from a fine scale
but idealized 3-D numerical hydrodynamic calculation of turbulent mix, for
example in an unstratified isochoric (constant and uniform density) fluid.

When this model is applied to an Eulerian calculation it is tacitly assumed
that an interface-preserving algorithm is used to prevent rapid numerical
diffusion of composition that may be entirely spurious (for example, if the
stratification is stable).
\section{Discussion}
Turbulent mixing is a problem of long standing, and many models have been
introduced.  Typically they involve several free parameters, and these may
be chosen to give excellent fits in regions of parameter space for which
appropriate data are available.  For example, the widely used BHR
model\cite{BHRZ92,UCHS01} contains ten fitting parameters.  These have been
chosen to fit a classic shock tube experiment\cite{ABMNPT82}.  With so many
parameters available the fit is excellent, but that does not establish its
validity when extrapolated outside the range of conditions to which they
were fitted.  For many important problems no data exist in the parameter
range of interest.  A simple model may be more robust than a sophisticated
one.

Any model, and especially one as oversimplified as that proposed here, needs to
be tested.  Quantitative experimental tests are few because of the difficulty of
measuring (and even defining in an experimentally accessible manner) effective
diffusion coefficients for complex heterogeneous flows.  Numerical simulation
may be a preferable approach, because there is no difficulty in defining spatial
distributions of composition, and even of fitting an effective $D$ to the
evolution of their means.  An unstable velocity field is set up on a fine mesh,
and the mixing calculated by the code is compared to that produced by the model
applied to a much coarser description of the mean velocity field.  Both the
validity of the model and the optimal value of any multiplicative coefficient
may be determined from this comparison.

Subgrid turbulent viscosity and heat transport present problems analogous to
that of subgrid turbulent mixing.  Sophisticated models exist, but also
contain several parameters (essentially closure parameters for the moment
equations).  It may be a reasonable rough approximation to use the model
proposed here for $D_{turb}$ and to take $\nu_{turb} = {\rm Sc}_{turb}
D_{turb}$, estimating Sc$_{turb}$ from numerical simulations like those used
to test the model for $D_{turb}$, or simply taking $\nu_{turb} = D_{turb}$
in the absence of such information.  Similarly, it may be reasonable to take
$\alpha_{turb} = D_{turb}$.

I thank G.~Dimonte, P.~E.~Dimotakis, R.~A.~Gore, L.~G.~Margolin,
D.~I.~Meiron and G.~B.~Zimmerman for useful discussions.

\end{document}